\documentstyle[draft]{npbproc}

\def\gtwid{\mathrel{\raise.3ex\hbox{$>$\kern-.75em\lower1ex\hbox{$\sim$}}}}
\def\ltwid{\mathrel{\raise.3ex\hbox{$<$\kern-.75em\lower1ex\hbox{$\sim$}}}}

\begin{document}
\title{ON THE ORIGIN OF FERMION MASSES \thanks{To appear in the
{\em Proceedings of the Workshop on Nonperturbative Properties of Chiral Gauge
Theories} (Rome, Mar. 1992), eds. L. Maiani, G. Rossi, and M. Testa.}}
\author{Robert E. Shrock%
  \thanks{Work partially supported by NSF grant PHY-91-08054.}}
\address{Institute for Theoretical Physics \\
         State University of New York \\
         Stony Brook, NY  11794-3840 \\
         USA}

\date{}

\runtitle{On the Origin of Fermion Masses}
\runauthor{R. E. Shrock}

\pubyear{1992}
\volume{25A}
\firstpage{1}
\lastpage{13}

\begin{abstract}
We review some recent work on nonperturbative properties of fermions and
connections with chiral gauge theories.  In particular, we consider one
of the ultimate goals of this program: the understanding of the actual
fermion mass spectrum.  It is pointed out that if quarks and leptons are
composite, their masses may be set by the physics of the preons and their
interactions in such a manner as to differ considerably from the Yukawa form
$m_f \propto v$ (where $v$ is the electroweak symmetry breaking scale) or
analogous forms involving $v$.
Some ideas of how this might work are given, and some
implications are discussed.
\end{abstract}

\maketitle

\section{INTRODUCTION}
One of the most important physical properties of a
chiral gauge theory is the fermion mass spectrum.  The standard electroweak
model can accomodate, but not predict, this spectrum.  The
understanding of fermion masses is one of the goals of nonperturbative studies
of quantum field theories.  Here we shall discuss several recent results
relevent to this subject.  In
particular, we shall offer some comments on the origin of fermion masses in
composite models.

\section{NONPERTURBATIVE PROPERTIES OF GAUGE-HIGGS-FERMION MODELS AND
 CONFINING MODELS OF WEAK INTERACTIONS}

\label{ghfw}

   Several lines of research have previously helped to give insights into
nonperturbative properties of fermion masses.   Among these was the
study of the (zero-temperature)
chiral transition and phase structure in lattice models with gauge, Higgs,
and fermion fields (for reviews, see \cite{rs87,rsrev}).
This work considered the SU(2) subgroup of the standard model (which can be
reexpressed as a vectorial theory), together with U(1) toy
models~\cite{osu,chid,u1km,wh,u1yuk,su2f,u1qq,su2yuk,ds,su2fd,ieq1}.  It
addressed the case where the bare fermion mass is zero, or, more generally,
very small compared with other mass scales in the theory.
Without Higgs fields, one knows that there is spontaneous chiral
symmetry breaking (S$\chi$SB) in such a theory, with dynamical generation of a
fermion condensate and fermion masses.  What happens when the
Higgs fields are added?

It was found that the theory consists of (i) a phase
which is analytically connected to the confinement phase of the pure
gauge-fermion theory in which S$\chi$SB continues to occur; and (ii) a phase in
the SU(2) theory (two phases in U(1) models) in which the chiral symmetry
is realized exactly, and the renormalized physical fermion mass is zero.
Starting from the gauge-Higgs theory, in which for fermions transforming
according to the fundamental representation of SU(2) (having
charge $q=1$ for U(1))
the confinement and Higgs phases are analytically connected, the addition of
fermions causes a dramatic change in the phase structure, since the
corresponding confinement and Higgs-like phases become distinct.
These studies made use of exact integrations over gauge fields
and analytic mean field theory methods in the limit of strong bare gauge
coupling limit,
together with numerical simulations with quenched and dynamical fermions.
The dependence upon fermion representations was studied in \cite{u1qq,ieq1}.

Among other applications, these results were applied to strongly interacting,
confining models of weak interactions~\cite{chid,su2f,su2fd,rs87}.  Such models
assume the existence of a confining phase in which, however, there is no
S$\chi$SB (which, given the rearrangement of the SU(2) theory in vectorial
form, would violate charge conservation).  The lattice studies found no
evidence for such a phase.
Two of these studies included Yukawa couplings~\cite{u1yuk,su2yuk}.

\section{LATTICE YUKAWA MODELS AND NONPERTURBATIVE UPPER BOUNDS ON FERMION
MASSES}

\label{ytriv}

   In another direction, a program was carried out to investigate the
nonperturbative properties of simple Yukawa models, again using lattice
regularization (for reviews, see \cite{jsrev,rsrev}).  Studies were carried out
using analytic methods and dynamical fermion simulations to determine the
phase structure of simple models with a real
scalar field, including the
 second-order phase transition lines where a continuum limit could be
 defined~\cite{js1,yr,yrn}.  The physical, renormalized Yukawa and scalar
 couplings $y_R$ and $\lambda_R$ were measured
as one approached such a continuum
limit.  In \cite{yr} these measurements
were carried out over a very large range
of bare Yukawa and scalar couplings $\lambda$ and $y$.
The data was consistent with the conclusion that both
$y_R$ and $\lambda_R$ would vanish as the lattice cutoff $a$ was taken to
zero, or equivalently, the ultraviolet cutoff $\propto 1/a$ was taken to
infinity, to construct the continuum theory.
Studies of different lattice
actions and models~\cite{rsrev,jsrev,js1,yr,yrn,ah,other,yra,lat90}
were carried out, and showed
that this conclusion was independent of the details of such actions, i.e.
was universal in the statistical mechanical sense.  These considered
both local lattice Yukawa interactions and another form which is natural for
staggered fermions, viz., Yukawa interactions defined on hypercubes.
(The behavior at strong bare Yukawa coupling, in contrast to that at weak bare
coupling,  was found to depend on the details of the lattice
action.)

   Reinterpreting the theory as an
effective one, with a physical cutoff in place (as with the standard
electroweak model, where this cutoff represents the onset of new physics),
one obtained nonzero values
for $y_R$ and $\lambda_R$. Given the requirement for the effective theory that
this cutoff must be larger than the characteristic masses in the theory, one
thus obtained upper bounds on $y_R$ and  $\lambda_R$.  In turn, these yielded
(nonperturbative)
upper bounds on the renormalized fermion and scalar masses.  It was found that
the renormalized fermion and scalar masses cannot be substantially larger than
the symmetry breaking scale $v$.

 The upper bound on the fermion mass generated
by a Yukawa interaction could provide a fundamental theoretical understanding
of an important feature of the observed fermion mass spectrum: the apparent
absence of any standard model fermions with masses substantially larger than
the electroweak symmetry breaking scale $v= 2^{-1/4}G_{F}^{-1/2} = 246$  GeV.
(Recall that the
latter absence is suggested by the LEP-SLC determination that there are three
generations of usual fermions with associated light or massless neutrinos,
together with phenomenological upper limits on $m_{top}$.)

    This work also constituted a generalization
of previous studies of the pure scalar sector and associated upper bounds
on the Higgs mass~\cite{phi4}) to the case
where fermions are included.  Note that
fermions might have produced a qualitative change in the behavior, since they
give rise to a negative term in the renormalization group equation
for the scalar coupling,
\begin{equation}
{{d\lambda} \over {dt}} = A_1 \lambda^2 + A_2 \lambda y^2 - A_3 y^4 \ ,
\qquad  A_{i} > 0
\end{equation}
where $t = \ln(q^2/\mu^2)$.
Measurements of renormalized couplings at other continuum limits at strong
bare Yukawa coupling were also carried out~\cite{other}.  Some results for the
phase structure of
gauged Yukawa models~\cite{u1yuk,su2yuk} and for renormalized couplings in a
global U(1) model~\cite{jssa} have also been obtained.

   One of the important results of this research was that in models with a
local Yukawa interaction, in addition to the symmetric paramagnetic PM1 phase
at small bare $y$, another symmetric phase was discovered at large bare $y$
\cite{yrn}.  Although the renormalized lattice
fermion mass $am_F$ was identically zero in the PM1 phase, it was nonzero
in the PM2 phase.  Hence, whereas $am_F$ vanished as one approached the usual
perturbative continuum limit along the FM-PM1 phase boundary at weak bare
$y$ from within the FM phase, it did not vanish when one approached the
nonperturbative continuum limit at the FM-PM2 phase boundary at strong bare
$y$.  This finding played a crucial role in the subsequent effort to
construct chiral lattice theories with the Wilson-Yukawa approach because it
meant that when one had another parameter $r$ with which to tune fermion
masses, one could satisfy at least one necessary condition of an acceptable
continuum limit: the doubler fermion modes would have $am_F$ finite as $a \to
0$, so that the actual doubler masses would diverge, while
the physical fermion mass could be rendered finite via the tuning.

\section{EFFORTS TO CONSTRUCT LATTICE CHIRAL GAUGE THEORIES}

       The present author has
been in one of the groups which studied an approach toward this goal using
the so-called Wilson-Yukawa term~\cite{cgt} for
decoupling fermion doubler modes.  To address some of the basic issues of
decoupling in numerical simulations in a simple context, it was
useful to consider subsectors of the standard model; we concentrated on the
U(1)$_Y$ hypercharge sector, since this
already embodies two features of the full
theory: (i) complex representations, and (ii) both left- and right-handed
fermions transforming as nonsinglets.  In contrast, (i) as is obvious from the
fact that SU(2) has only real representations, the SU(2) sector, taken
by itself (with an even number of left-handed doublets to avoid the Witten
anomaly), can be reexpressed as a vectorial theory, a fact used in our earlier
work mentioned in section \ref{ghfw}; and (ii) in the SU(2) sector by itself,
the right-handed fermions are singlets.
Analytic hopping parameter expansions for the full gauge theory and
numerical simulations of the global limit of this theory were carried out and
showed that it is possible to decouple fermion doublers and tune renormalized
lattice fermion masses $am_F \to 0$ as $a \to 0$ to obtain finite fermion
masses in the continuum limit~\cite{alss,als}.  Similar results were obtained
by other authors for SU(2)
using the Wilson-Yukawa approach (\cite{acam}; for a review,
see \cite{jsrev}).  Formulations for the full SU(2) $\times$ U(1)$_Y$
theory have also been considered~\cite{ihl,smit}, and a discussion of
decoupling has been given~\cite{su2xu1}.

   There have, however, been several problems with the Wilson-Yukawa and other
approaches to constructing chiral gauge theories on the lattice.  First, the
fermion doublers do not decouple perturbatively; nonperturbative behavior is
crucial to achieve this decoupling.  Because of this, it is not easy to make
contact with the perturbative calculations which have been quite successful in
the continuum electroweak theory.  Second, the functional integral does
not have a Gibbs measure and there is no known reliable method for numerical
simulations of such a theory.  This is a severe drawback
since it means that numerical simulations are limited to the quenched
approximation and hence are clearly missing important physics, since they
are not even sensitive to whether a chiral gauge theory is anomaly-free or is
anomalous.  Third, the formulation does not
adequately deal with instanton-induced breaking of global symmetries, which
occurs for the full electroweak theory, although not for a U(1) gauge
group~\cite{inst,eich,banks}.
The question of whether the Wilson-Yukawa
formulation yields reasonable behavior for fermions in the symmetric
strong-coupling phase and the continuum limit from the broken phase
has also been investigated recently~\cite{gpsbds,u1ns,salat91,psr}.

    There are other approaches besides the one based on the
Wilson-Yukawa term, such as that with mirror fermions~\cite{mont} and
that of the Rome group~\cite{rome}.  It is hoped that as the
connections between these different approaches are clarified one may be able
to overcome some of the current problems and make further progress.

\section{COMPOSITENESS AND THE ORIGIN OF FERMION MASS}

\subsection{General}

   The work above has yielded insights into the nonperturbative behavior of
fermions.  Indeed, the nonperturbative upper bound on fermion masses
discussed in section
\ref{ytriv} may help to explain why there are apparently no
standard model fermions with Yukawa-generated masses $m_f$ much larger than
the EWSB scale $v$.  However,
this is far from actually calculating the fermion mass spectrum.
The same is true of current efforts to construct lattice chiral gauge
theories, even if some of the present difficulties can be overcome.

   In view of this, it seems worthwhile to step back and reconsider some
basic aspects of fermion mass generation.
Granting that the standard model cannot predict the fermion masses
quantitatively, one may ask a more modest question: does it at least give a
rough, qualitative understanding of these masses.  For this purpose, let us
consider the W and Z masses for comparison.  Assume that one had not measured
$\sin^2 \theta_W$.  Using the known value of $v$, and a guess
$|\sin \theta_W| = 0.5$, midway between its limits, one would obtain a
tree-level result $m_W = 75$ GeV, reasonably close to the actual W mass.
However, if one does something similar for a fermion mass, using the tree-level
formula $m_f = hv/\sqrt 2$ where $h$ is a generic Yukawa coupling, and taking
$h$ to be the size of a typical small coupling like $e = 0.3$ (and
neglecting fermion mixing), one gets $m_f \simeq 50$ GeV, which is too large
by a factor of $10^5$ for $e$ and about $10^4$ for $u$ and $d$.
In the context of the standard model, one must simply accept these
extremely small physical Yukawa couplings for the charged leptons and all of
the quarks except for top, and relegate any attempt to explain them to new
physics beyond this model.

\subsection{A Hypothetical World}

   Here we would like to investigate the origin of fermion masses within the
context of composite models~\cite{fm}.
It is instructive first to consider an example
which is pedagogical but makes no pretense of being realistic.  Imagine a world
in which physicists knew about the electron, proton, and neutron (but not the
muon, pion, or heavier particles) and had
inferred the existence of the $\nu_e$ from nuclear and neutron
beta decay.  Assume that they had developed the theory of quantum
electrodynamics and the associated ideas of renormalizability and local gauge
invariance.  Assume further that they had not measured the magnetic moments
$\mu_p$ or $\mu_n$ and that, to within the accuracy of their measurements, the
axial vector coupling $g_A \simeq 1$.  Finally, assume that other data was
consistent with the $p$ and $n$ being pointlike particles.  These physicists
could well have constructed an SU(2) $\times$ U(1)$_Y$ electroweak
gauge theory,
with the (first-generation) leptons assigned in the usual way and the nucleons
in an SU(2) doublet
\begin{equation}
\psi_{N,L} = { p \choose n}_L  \label{nrep}
\end{equation}
with weak hypercharge $Y=1$ and right-handed SU(2) singlets $p_R$ and $n_R$
with $Y=2$ and 0, respectively.  They would observe with contentment that this
theory is anomaly-free.  From their measurements of nuclear beta decay, they
would know an approximate value for $G_F$ and hence $v$, and, with $g \sim e$,
they would thus know that $m_W$ and $m_Z$ were far above the energy scale
reached by their data, so that it was consistent that they had not yet observed
these particles.

   The key point concerns how these physicists would try to
 explain the origin of the
nucleon mass.  They would realize that bare nucleon mass terms
$m_p (\bar p_L p_R + h.c.) + m_n(\bar n_L n_R + h.c.)$
would violate the electroweak gauge symmetry.  Consider first the situation if
the physicists had incorporated the usual Higgs doublet
$\phi = {\phi^+ \choose \phi^0}$ in their theory, with $<\phi^0> = v/\sqrt
2$ and
\begin{equation}
\phi^0 - <\phi^0> = {{(H + i\chi)} \over \sqrt 2}
\end{equation}
Given that they applied the
constraints of renormalizability and exact gauge invariance in their
electroweak  Lagrangian,
it is likely that they would thus attribute the nucleon mass generation to a
Yukawa interaction of the form
\begin{equation}
-{\cal L}_{Y,N} = h_n \bar\psi_{N,L} \phi n_R + h_p \bar p_R
\phi^{\dagger} \psi_{N,L} + h.c.
\label{nyuk}
\end{equation}
They would thus predict that
\begin{equation}
m_{p,n} \propto v \label{myuk}
\end{equation}
If they avoided the Higgs mechanism and instead assumed a dynamical symmetry
breaking scheme like extended technicolor (ETC), they would predict that
\begin{equation}
m_N \sim {\Lambda_{TC}^3 \over \Lambda_{ETC}^2} \propto {v^3 \over
\Lambda_{ETC}^2} \label{metc}
\end{equation}
In both cases, they would expect the nucleon mass to be set by the electroweak
symmetry breaking scale $v$.  In both cases, they would be wrong.
In fact, the nucleon mass arises
from its compositeness and the associated QCD binding of the (light)
quarks of which it is composed, and is
\begin{equation}
m_N \sim const. \times f_{\pi} \sim const. \times \Lambda_{QCD}
\label{mqcd}
\end{equation}
(where the constants are somewhat larger than unity).

   These physicists' logic was correct, except for the falsity
of their initial
assumption that the nucleons were pointlike.  We suggest that, in a similar
manner, the conventional view of the masses of the quarks and the charged
leptons and (left-handed) neutrinos as
arising in a manner closely connected with the EWSB scale $v$ may need revision
if these particles are, in fact, composite.

    Of course, the compositeness of the nucleon has
fundamentally different properties from the possible compositeness of quarks
and leptons.  In models of quark and lepton
substructure, one must avoid the spontaneous chiral symmetry breaking which
occurs in QCD, since this would yield composite fermion masses naturally of
order the inverse compositeness scale,
$\Lambda_c = R_{c}^{-1} \gtwid 1$ TeV.
It is also important to note that the S$\chi$SB in QCD and associated quark
condensate $<\bar q_L q_R + h.c.>$ (spontaneously)
breaks electroweak gauge invariance; this was, indeed, the original
inspiration for technicolor.  However, this source of EWSB is much smaller
than the dominant source: $f_{\pi}/v \simeq 4 \times 10^{-4}$.
Consequently, the nucleon mass $m_N$ is largely
independent of the (dominant) scale of EWSB, $v = 250$ GeV, even though
it is set by a phenomenon which breaks electroweak symmetry.

    It is interesting to consider what would happen if one changed
the EWSB scale $v$ while keeping
$\Lambda_{QCD}$ fixed.  (One might object that in a complete theory, this would
be a logical impossibility, since both would be uniquely predicted.  However,
we believe that such hypothetical changes of various quantities can give useful
insights; for example, large-$N_c$ methods have been of value in understanding
QCD even though it
may well be true that in a complete theory it would, strictly speaking, not be
possible to change $N_c$.)
 In the hypothetical world discussed in \cite{fm}, it was postulated
that the physicists had not observed the pions, since if they had, they would
probably have realized that the nucleons are not pointlike (and also would have
discovered the muon) and would not have
tried to construct an SU(2) $\times$ U(1)$_Y$ electroweak theory with the
assignment (\ref{nrep}).
It follows that if one were to increase $v$ up or down
by a factor of 100, say, although the strength of weak interactions would
decrease or increase by a factor of $10^4$
relative to electromagnetic and strong
interactions, the nucleon mass would not change significantly.  Of course, one
may feel that the exclusion of the pion makes this hypothetical world too
unrealistic.  In an appendix we briefly comment on what might happen in our
actual world if we again imagined changing the ratio $v/\Lambda_{QCD}$.

\subsection{An Illustrative Model}

   Let us proceed to discuss some ideas for mass generation in composite models
of quarks and leptons.  As is well known, such models must confront the problem
of obtaining a composite fermion mass $m_{cf}$ which is much smaller than the
scale of compositeness, $\Lambda_c$. As noted above, this
requires that there be no spontaneous chiral symmetry breaking by the
interaction which binds the preons.  There is by
now some understanding of how one might obtain massless composite fermions
starting from a preonic gauge theory with massless preons, as the result of
unbroken global chiral
symmetries \cite{thooft,drs,pw,bars,marsh,largen} (for a review and further
references, see \cite{pesk}).

   Thus, let us consider a
preonic chiral gauge theory with a set of massless fermionic preons
transforming according to
complex representations of the gauge group $G$, together with another set
transforming according to real representations.  We assume that (i) the
preon representations are restricted so that the theory is asymptotically
free; and (ii) it is valid to view the theory as confining, so that the
spectrum consists of singlets under $G$.  It  might seem more natural
to expect that the the chiral gauge invariance would be
spontaneously broken by the formation of a bilinear chiral preonic
condensate somewhat analogously to the situation in technicolor,
but in some cases there may be a kind of
complementarity between the confinement and broken or Higgs-like
pictures~\cite{drs,tum}.
(Admittedly, complications with this idea have been found
in specific lattice studies~\cite{rs87,chid,su2f,su2fd,rs85}.)
Further, assume that the requisite anomaly matching consistency
conditions~\cite{thooft,drs} are satisfied so that the
global chiral flavor symmetries of this theory lead to massless composite
fermions.  Some of the preons
must be electroweak nonsinglets; let us assume that there are also
some electroweak-singlet preons and that some of these transform according to
real representations of $G$.

Now we address the issue of giving masses to the hitherto massless composite
fermions.   We specialize to composite fermions which contain at least one
real-representation, electroweak-singlet preon.  A conjecture is that we may be
able to produce
masses for these composite fermions by letting the mass(es)
of the constituent real-representation electroweak-singlet preon(s) be
nonzero.  The actual calculation of the mass of the
composite fermion including its dependence upon the mass(es) of its
constituent
real-representation electroweak-singlet preon(s) is a difficult task which
one is not able to carry out at present.  In particular, it is not known
whether giving a preon a mass $m_{pr} << \Lambda_c$ will immediately render
the composite fermion mass nonzero (the strong form of 't Hooft's decoupling
condition \cite{thooft}, sometimes called the ``persistent mass condition'')
or whether $m_{pr}$ must exceed some critical
value $m_{cr}$ before the composite fermion mass becomes nonzero \cite{pw}
(denoted here as the ``suppressed composite mass" or SCM case).
Alternatively, an electroweak-nonsinglet composite fermion containing an
electroweak-singlet preon might unbind and disappear from the spectrum at the
point where it would otherwise have gained a nonzero mass.  The possibility
considered here,  that the electroweak-nonsinglet composite fermion
could pick up a mass while staying bound,
could work in either the case of persistent mass or suppressed composite
mass behavior, although it would probably
operate more naturally in the SCM case.  Let us consider first the case where
the persistent mass behavior applies, i.e. $m_{pr} > 0 \Rightarrow m_{cf} >
0$. For small $m_{pr}$, we may heuristically write
\begin{equation}
m_{cf} \sim \Lambda_c \Bigl ({m_{pr} \over \Lambda_c} \Bigr )^\beta
\label{pmc}
\end{equation}
If $\beta < 1$, then $m_{cf}$ would rise rapidly, with initially
infinite slope, as $m_{pr}$ increases from 0, so
that it would not be natural for $m_{cf}$ to be $<< \Lambda_c$.  However,
$\beta$ could also be $> 1$; in this case, $m_{cf}$ would rise with
initially zero slope as
$m_{pr}$ increases from zero, and hence in this case, it
might be possible to satisfy the requisite condition that
$m_{cf} << \Lambda_c$ even with $m_{pr}$ not $<< \Lambda_c$.
Analogously, for the case of SCM behavior, for small $(m_{pr} -
m_{cr})/\Lambda_c$, one may heuristically write
\begin{equation}
m_{cf} \sim \Lambda_c \Bigl ({m_{pr} \over m_{cr}} - 1 \Bigr )^\beta
\label{scm}
\end{equation}
where $\beta$ plays the role of a critical exponent analogous to the critical
exponent for the order parameter in statistical mechanics.
The actual critical behavior may of course differ from the simple algebraic
 form in eq. (\ref{scm}); for
example, there could be an additional logarithmic factor, which would not be
important here, or the critical singularity could be an essential zero, which
would manifest itself as $\beta = \infty$.  As before, if $\beta < 1$ ($> 1$),
then $dm_{cf}/dm_{pr} = \infty$ (0) respectively as
$m_{pr} - m_{cr} \to 0^+$.
In the case $\beta > 1$, one could naturally obtain $m_{cf} << \Lambda_c$.
Parenthetically, it might be noted that although $\beta$ is typically $< 1$ in
statistical mechanical models, there do exist cases where $\beta > 1$.  For
example, in the exact solution of the large-$N$ O(N) nonlinear sigma model in
$d=2+\epsilon$ dimensions \cite{bls}, the critical exponent for the mass gap,
analogous to $\beta$ here, is $1/\epsilon$ and $\to \infty$ as $d \searrow 2$.
It is also possible that the nonzero preon mass $m_{pr}$ might be
$<< \Lambda_c$.
A priori, one might regard this as contrived, but one may recall that in the
case of hadrons, the first generations of quarks have
masses which are much smaller than the compositeness scale.

If, indeed, the electroweak-nonsinglet composite fermion remains bound and
does pick up a mass, this phenomenon has the striking feature that this mass
arises
because of the nonzero mass(es) of the constituent
electroweak-singlet preon(s), and the latter mass(es) do not violate
electroweak symmetry.  It is in this sense that one may interpret this by
saying that
the composite fermion mass arises in a manner which is largely
independent of electroweak symmetry breaking.  As was noted in
\cite{fm}, we say ``largely'' because there will, of course, be electroweak
radiative corrections to the composite fermion mass which
bring in a connection to EWSB, but
these are formally of order $\alpha$ and may be a small effect.

   Needless to say, at this stage this is just a speculative idea.
In order to demonstrate conclusively that the idea can be
implemented or is impossible, one would need a realistic model of quark and
lepton compositeness (in which one could carry out
reliable calculations).  Unfortunately, such a model does not exist.   One
already knows that the properties of such a model would have to be quite
different from those of any hitherto understood type of compositeness.
The issue of the dependence of the composite
fermion mass on the masses of its constituent preons illustrates the
sort of counterintuitive behavior that might happen; {\it a priori},
it is natural to expect that as one increased the mass of some preon from zero
to a small nonzero value, the previously
massless composite fermion containing this preon would immediately pick up a
nonzero mass.  It would appear unnatural to think that the interaction
responsible for preonic binding could increase the binding energy in
just such a manner as to compensate exactly for
the increase in $m_{pr}$ so that the composite fermion mass $m_{cf}$ remained
fixed at zero as $m_{pr}$ changed.
But this behavior cannot be ruled out and may, indeed,
occur~\cite{pw}.  In a similar manner, it may seem counterintuitive that an
electroweak-nonsinglet composite fermion could pick up a mass from a
constituent electroweak-singlet preon.  Of course, the third type of behavior
may also happen: as one
turns on the mass $m_{pr}$ of the electroweak-singlet preon
(or makes it larger than
the value $m_{cr}$ in the SCM picture), all electroweak nonsinglet composite
fermions which contain this preon may unbind.

\subsection{Discussion}

   One of the appealing aspects of the idea discussed here is that it might
ultimately  be one way to help explain (more exactly, to remove) the
old mystery in the standard model of why the
known fermion masses except for $m_{top}$ are much smaller than the
electroweak symmetry breaking scale.
Perhaps this mystery arose because of the
overly restrictive assumption that the quarks and leptons are pointlike, and
the resultant assumption that their masses arise from a Yukawa coupling, with
$m_f \sim v$.  If in fact these masses are dominantly set by a scale other than
the EWSB scale $v$, then the above feature of the fermion mass spectrum is no
longer necessarily a mystery.  Of course, this is not the only way to explain
$m_f << v$.  Indeed, in the original form of extended technicolor,
one had the opposite problem that it was difficult to get sufficiently large
fermion masses, and in walking technicolor, it may be possible to obtain a
realistic fermion spectrum while at the same time obeying constraints from
precision electroweak measurements~\cite{ta}.

   One may ask whether in our approach it is not a mystery that fermion masses
are not $>> v$.  To answer this, let us consider another {\it gedanken}
world in
which there is a large mass range between $v$ and $\Lambda_c$: $v <<
\Lambda_c$. (Recall that present limits on quark and lepton compositeness
\cite{pdg} only put $\Lambda_c \gtwid 1$ TeV, which is not $>> v$.)
Now consider an effective field theory defined at a reference energy
$E_{ref.}$ such that
\begin{displaymath}
v << E_{ref.} << \Lambda_c
\end{displaymath}
Since $E_{ref.} << \Lambda_c$, the
fermions would appear pointlike up to small corrections, and since
 $E_{ref.} >> v$, the electroweak
symmetry would be exact to order $O(v/E_{ref.})$.
The usual argument that a mass term for an electroweak nonsinglet fermion is
approximately forbidden by the approximate electroweak symmetry would hold
and would imply
that such a mass must be zero, at least up to $O(v/E_{ref.})$ times the
reference energy scale $E_{ref.}$, i.e. up to terms $\ltwid v$.
This constraint is of course satisfied by
the known fermion masses, but would rule out $m_f >> v$.  Hence, within our
context, it is understandable that fermion masses are not large compared with
$v$.

   One may also ask what would happen at finite temperature.  First, we note
that the `t Hooft anomaly matching constraints were only formulated at
zero temperature, and there are complications in trying to apply them at finite
temperature.  For example, it was found not to be possible to conclude that
(for $N_f \ge 3$ flavors, where these constraints
cannot be realized in the symmetric mode)
they imply a zero-mass goldstone pion at finite temperature $T < T_{deconf.}$
in QCD~\cite{ito}, in contrast to the $T=0$ case where they do.
Hence, it is not clear that one can use these constraints to infer conclusions
about the zero-mass fermion content of composite models at finite $T$ in the
manner that one does at $T=0$.  It is true that one could, heuristically at
least, repeat the sort of effective field theory analysis above at finite $T$.
Consider a temperature $T_{ref.}$ above the EWSB scale $v$ but below the
compositeness scale $\Lambda_c$.  Then again the composite fermions appear
pointlike up to small corrections.  An old analysis in terms of the
temperature-dependent effective potential~\cite{linde} implies that the theory
would have undergone a phase transition at $T = T_{cr} \simeq v$ and at
$T > T_{cr}$ the
electroweak symmetry would be exact.  This analysis is based on a perturbative
treatment of the Higgs potential in the standard model Lagrangian.  There might
also be nonperturbative effects such as EWSB fermion condensates associated
 with dynamical symmetry breaking, but these would presumably also also vanish
identically for $T > T_{cr}$.  If the electroweak symmetry is really exact at
$T > T_{cr}$ (and not just exact up to small terms of order $O(v/T)$),
then this implies the absence of any massive electroweak-nonsinglet
fermions in the spectrum.  The difference with the previous discussion would be
that for $T=0$,
at $E_{ref.}$, the EW symmetry is only exact up to small terms of order
$O(v/E_{ref.})$, whereas for $T > T_{cr} \simeq v$ it is truly exact.  If
there were EW-nonsinglet composite fermions with masses produced by constituent
EW-singlet preons, they would evidently have to unbind and disappear from
the spectrum as $T$ increased past $T_{cr}$.

    Aside from the absence of realistic composite models and the puzzle of
 preon dynamics, one unappealing feature of this approach
is that it pushes the explanation of quark and lepton masses down to
the next level down in structure, to the nonzero preon masses.  One must ask
what is the origin of these preon masses.  Of course, we have already
encountered a similar situation before; for example, the properties of
chemistry are determined by quantities such as the size of $\alpha$ and
$m_e/m_N$ which are not explained at that level.
Moreover, one may ask why the
preons themselves are taken to be pointlike.  Perhaps they too are composite
(where might the successive levels of substructure end?).

\section{Conclusions}

    Let us conclude with some implications relevant to the ongoing effort to
construct chiral gauge theories on the lattice and to understand their
nonperturbative behavior.  There is a strong concentration on trying to put the
standard electroweak model and simplifications thereof (e.g. to globally
symmetric models) on the lattice.  But the standard model
cannot explain the observed fermion masses.  Since these are one of main
remaining mysteries to be understood, we believe it would be worthwhile to
expand current efforts to (even toy) models which go beyond the
standard model, such as those which do not involve fundamental Higgs, but
rather some form of dynamical symmetry breaking.
Nonperturbative results on Yukawa models such as
the observation, in a simple Z(2) model~\cite{yrn}, of the symmetric PM2
phase at large bare $y$ where the renormalized fermion mass
is nonzero even though the Z(2) symmetry is exact,
have already shown that the actual behavior of fermions can
differ strongly with that in perturbation theory.  The lattice studies
supporting the conclusion that Yukawa theories are non-interacting in the
continuum limit also show how true behavior can differ strongly from naive
perturbation theory.

    It would be desirable to go
further and try to construct chiral theories in which one can investigate
compositeness, using, e.g. lattice methods.  Thus one would like to formulate a
chiral preonic gauge theory which might underlie the standard model.  Needless
to say, there appear formidable obstacles to a quantitative study of such a
theory on the lattice.  Aside from the obvious problems about decoupling
fermion doublers and matching onto electroweak perturbation theory, it is
difficult to see how one can deal with widely disparate mass scales like $m_e$,
$v$, and $\Lambda_c$ in lattice studies.  Part of the success of lattice QCD
relied on the fact that the QCD scale, $\Lambda_{QCD}$ was comparable to the
other physical mass scales such as $K^{1/2}$ ($K$ = string tension),
$m_{\rho}$,
$m_N$, and $m_{glueball}$.  A lattice formulation of even the standard model,
let alone models of dynamical symmetry breaking or composite quarks and
leptons, would have to contend with very different mass scales.  This is a
fitting challenge for the future.

\begin{acknowledge}
We thank G. Maiani, G. Rossi, and M. Testa for their warm hospitality during
this stimulating workshop.
\end{acknowledge}

\bigskip
\leftline{APPENDIX}
\bigskip

    In the hypothetical example  in Ref. \cite{fm}, it was postulated
that the physicists had not yet discovered the pion, since if they had, it is
unlikely that they would have tried to construct the SU(2) $\times$ U(1)$_Y$
electroweak gauge theory with the particle
assignment in (\ref{nrep}).  However, it is
of interest to examine a different hypothetical question: what would happen to
the nucleon mass in the real world if one changed the EWSB scale $v$, while
 keeping the QCD scale $\Lambda_{QCD}$ and hence also $f_{\pi}$ essentially
fixed?  We shall
offer some brief comments on this here.  For simplicity, we shall restrict to
first-generation fermions.
First, consider the limit where there are no electroweak
interactions and $m_u = m_d =0$. Then the physics is invariant under the
global chiral symmetry
$G_{\chi} =$ SU(2)$_L \times$ SU(2)$_R$.  The spontaneous breaking of this
symmetry has by now been understood at a fundamental level in terms of QCD, and
the associated parameter $f_{\pi}$ approximately calculated in terms of
$\Lambda_{QCD}$ via lattice methods.  However, because of its
 simplicity, the old
phenomenological description provided by the sigma model is still of use.  We
recall that the linear sigma model has the interaction
$$
{\cal L}_{L \sigma M} = g_{\pi NN}\bar N( \sigma + i \gamma_5
\pi \cdot \tau)N \eqno(A.1)$$
where $N = {p \choose n}$.
The vev $<\sigma> = f_{\pi}$ simulates the actual S$\chi$SB of $G_{\chi}$ by
QCD.  (Since there is probably no resonance which can be reasonably
identified as the $\sigma$ and is sufficiently narrow to be considered a
particle~\cite{pdg}, one might prefer the nonlinear sigma model.  Since the
latter involves a nonpolynomial Lagrangian, we shall restrict ourselves here to
the somewhat simpler linear sigma model. Because of the strong coupling
$g_{\pi NN}$, one cannot use perturbation theory with ${\cal L}_{L \sigma M}$,
but if one were to ignore
this, the nucleon mass would be given by $m_N = g_{\pi NN}f_{\pi}$, and this
can be used as a rough guide to the actual value of $m_N$.  (For comparison, in
a different
effective Lagrangian approach involving skyrmions, where the nucleon appears as
a soliton of the field $U= \exp(i \pi \cdot  \tau/f_{\pi})$, one
obtains~\cite{witt} $m_N \simeq f_{\pi}( 73/g_{\rho} + 0.007 g_{\rho}^3)$.)

   Now let us turn on the electroweak interactions.
These explicitly break $G_{\chi}$, but this is a small effect, of order
$\alpha$.
Let us denote the electroweak Nambu-Goldstone modes in the
 fictitious absence of the hadronic
sector as $w^{\pm}$ and $z$.  (If one chooses to work in the Higgs picture,
then $w^{\pm} = \phi^{\pm}$ and $z = \chi$.)
These will mix with the almost goldstone modes in the hadronic sector
according to
$$
\left( \matrix{ w^{+}' \cr \pi^{+}' \cr} \right ) = R(\theta_+)
\left( \matrix{ w^+ \cr \pi^+ \cr} \right )  \eqno(A.2)
$$
and
$$
\left( \matrix{ z' \cr \pi^{0}' \cr} \right ) = R(\theta_0)
\left( \matrix{ z \cr \pi^0 \cr} \right ) \eqno(A.3)
$$
where
$$R(\theta) = \left( \matrix{ \cos \theta & \sin \theta \cr
                           -\sin \theta  & \cos \theta \cr } \right )
\eqno(A.4)
$$
The $w^{\pm}'$ and $z'$ denote the combinations which are absorbed to form
the longitudinal components of the W$^\pm$ and Z bosons, and the
$\pi^{\pm}'$ and $\pi^{0}'$ are the orthogonal states.
If one works in the Higgs picture and the linear sigma model, then
there will also be mixing among $H$ and $\sigma$:
$$
\left( \matrix{ H' \cr \sigma' \cr} \right ) = R(\theta_s)
\left( \matrix{ H \cr \sigma \cr} \right ) \eqno(A.5)
$$
The mixing angles may be calculated by diagonalizing the mass matrices in each
sector.  For example, assume that one includes $H$ and $\sigma$; then the
corresponding mass matrix is
$$
\left( \matrix{ m_{H}^2  & m_{H \sigma}^2 \cr
                  m_{H \sigma}^2  & m_{\sigma}^2 \cr} \right ) \eqno(A.6)
$$
The resultant mixing angle is given by
$$
\tan 2 \theta_s = {{2m_{H \sigma}^2} \over {m_{H}^2 - m_{\sigma}^2}}
\eqno(A.7)
$$
The term $m_{H\sigma}^2$ represents the coupling between the electroweak and
hadronic sectors and is presumably of order $(\alpha/\pi)f_{\pi}^2$, similar to
estimated electroweak corrections to meson masses squared.  Using the
expectation $m_H \sim v$ and the input $m_{\sigma} \sim const. \times f_{\pi}$
in the context of the sigma model, (A.7) reduces to
$$
\theta_s \simeq {\alpha \over \pi}\Bigl ({f_{\pi} \over v}\Bigr )^2
 \qquad  (\hbox{for} \ f_{\pi} << v)
\eqno(A.8)
$$
Evaluating this gives
$$
\theta_s \simeq O(10^{-9})  \eqno(A.9)
$$
(Putting in the nonzero quark masses $m_u, m_d << f_{\pi}$ will not alter this
significantly.)
This mixing measures how the electroweak sector would alter the composition of
the $\sigma$ field and is evidently quite small.  From (A.8) it
is clear that changing $v$ up or down by
even a large factor would not induce significant mixing as long as $v >>
\Lambda_{QCD}$.
The electroweak interactions would also affect the size of
$<\sigma>$.  It is plausible that the shift in $<\sigma>^2$ might be an order
$\alpha$ fractional effect, like the electroweak
 mass shifts of meson masses squared or the $n-p$ mass difference.
(Note that propagator correction graphs involving virtual W or Z lines
have dominant terms of roughly the same order as pure QED graphs; see, e.g.,
\cite{bwl}.)    Assuming this is the
case, the nucleon mass would only shift slightly due to the electroweak
 interactions,
$$
m_N = a_0 (f_{\pi})_0\Bigl [ 1 + O\Bigl ({\alpha \over \pi}\Bigr ) \Bigr ]
\eqno(A.10)
$$
where $a_0$ and $(f_{\pi})_0$ refer to the values of the constant in
(\ref{mqcd}) when electroweak interactions are turned off.  This was the
 origin of the comment in \cite{fm} that although electroweak radiative
corrections bring in a
connection with the EWSB scale, they may be a small effect.  One may also try
to approach this issue by writing down a guess for cross-terms in a combined
EW-sigma model potential for $\phi$ and $(\sigma, {\bf \pi})$.  Here there is
some indication of apparent hierarchy problem involving the vev's $<\sigma>$
and $<\phi>$.   Since we know the answer from nature, that $<\sigma>$ is
$<< v$, it seems likely to us that the apparent hierarchy problem is an
artifact of the phenomenological potential approach and in particular, the
perturbative treatment of the $\sigma$ model Lagrangian despite the strong
coupling $g_{\pi NN}$.   Finally, we note that clearly
 if one were to reduce the
EWSB scale to the point where $v \simeq \Lambda_{QCD}$, then the
mixing angles in (A.2), (A.30, and (A.5) would be large, and the
strength of the weak interaction
would be comparable to that of the electromagnetic
interaction.  The range of variation of $v$ considered in \cite{fm} was
restricted to preclude this possibility and keep things relative simple.

\end{document}